\documentclass[preprint,5p,times,twocolumn]{elsarticle}
\usepackage{amssymb}
\usepackage{amsmath}
\usepackage{epsfig}
\journal{Phys lett B}
\begin{document}
\begin{frontmatter}
\title{Analytical Solution for the SU(2) Hedgehog\\ Skyrmion and Static
Properties of Nucleons}
\author{Duo-Jie Jia$^{\ast}$}
\ead{jiadj@nwnu.edu.cn}
\author{Xiao-Wei Wang}
\author{Feng Liu}
\cortext[ast]{Corresponding author}
\address{Institute of Theoretical Physics, College of Physics
and Electronic \\Engineering, Northwest Normal University,\textit{\
Lanzhou 730070, P.R. China}}
\begin{abstract}
An analytical solution for symmetric Skyrmion was proposed for the
SU(2) Skyrme model, which take the form of the hybrid form of a
kink-like solution and that given by the instanton method. The
static properties of nucleons was then computed within the framework
of collective quantization of the Skyrme model, with a good
agreement with that given by the exact numeric solution. The
comparisons with the previous results as well as the experimental
values are also given.
\end{abstract}
\end{frontmatter}
\section{Introduction}
The Skyrme model \cite{Skyrme} is an effective field theory of
mesons and baryons in which baryons arise as topological soliton
solutions, known as Skyrmions. The model is based on the pre-QCD
nonlinear $\sigma $ model of the pion meson and was usually regarded
to be consistent with the low-energy limit of large-N
QCD\cite{Witten}. For this reason, among others, it has
been extensively revisited in recent years \cite%
{ANW,JR,Sutcliffe,Battye97,Atiyah89,Battye} (see,
\cite{Zahed,Topsoliton}, for a review). Owing to the high
nonlinearity, the solution to the Skyrme model was mainly studied
through the numerical approach. It is worthwhile, however, to seek
the analytic solutions \cite{JMP,Ponciano,Yamashita} of Skyrmions
due to its various applications in baryon phenomenology. One of
noticeable analytic method for studying the Skyrmion solutions is
the instanton approach proposed by Atiyah and Manton\cite{Atiyah89}
which approximates critical points of the Skyrme energy functional.

In this Letter, we address the static solution of hedgehog Skyrmion in the $%
SU(2)$ Skyrme model without pion mass term and propose an analytical
solution for the hedgehog Skyrmion by writing it as the hybrid form
of a kink-like solution and the analytic solution obtained by the
instanton method \cite{Atiyah89}. Two lowest order of Pad\'{e}
approximations was used and the corresponding solutions for Skyrmion
profile are given explicitly by using the downhill simplex method.
The Skyrmion mass and static properties of nucleon as well as delta
was computed and compared to the previous results.
\section{Analytic solution to Skyrme model}
The $SU(2)$ Skyrme\ action \cite{Skyrme} without pion mass term is
given by\\
\begin{equation}
\mathcal{S}^{SK}=\int d^{3}x\left[ -\frac{f{}_{\pi}^{2}}{4}Tr(L_{\mu })^{2}+\frac{%
1}{32e^{2}}Tr([L_{\mu },L_{\nu }]^{2})\right]  \label{SK}
\end{equation}
in which $L_{\mu }=U^{\dag }\partial _{\mu }U$, $U(x,t)\in SU(2)$ is
the nonlinear realization of the chiral field describing the $\sigma
$ field and $\pi $ mesons with the unitary constrain $U^{\dag }U=1$,
$2f_{\pi }$ the pion decay constant, and $e$ a dimensionless
constant characterizing nonlinear coupling. The Cauchy-Schwartz inequality for (\ref{SK}) implies%
\cite{Faddeev} $E^{SK}\geq 6\pi ^{2}(f_{\pi }/e)|B|$, where $B\equiv $ $%
(1/24\pi ^{2})\int d^{3}x\varepsilon ^{ijk}Tr(L_{i}L_{j}L_{k})$ is
the topological charge, known as baryon number. Using the hedgehog ansatz, $%
U(x)=\cos (F)-i(\hat{x}\cdot \vec{\sigma})\sin (F)$ ($\vec{\sigma}$
are the three Pauli matrices) with $F\equiv F(r)$ depending merely
on the radial coordinate $r$, the static energy for (\ref{SK})
becomes
\begin{equation}
E^{SK}=2\pi (\frac{f_{\pi }}{e})\int dx\left[ x^{2}F_{x}^{2}+2\sin
^{2}(F)(1+F_{x}^{2})+\frac{\sin ^{4}(F)}{x^{2}}\right] \label{Eden}
\end{equation}
with $x=ef_{\pi }r$ a dimensionless variable and $F_{x}\equiv
dF(x)/dx$. The equation of motion of (\ref{Eden}) is%
\begin{equation}
\left( 1+2\frac{\sin ^{2}F}{x^{2}}\right)
F_{xx}+\frac{2}{x}F_{x}+\frac{\sin (2F)}{x^{2}}\left(
F_{x}^{2}-1-\frac{\sin ^{2}F}{x^{2}}\right) =0, \label{EOM1}
\end{equation}
where the boundary condition $F(0)=\pi ,F(\infty )=0$ will be
imposed so that it corresponds to the physical vacuum for
(\ref{SK}):$U=\pm 1$. The equation (\ref{EOM1}) is usually solved
numerically due to its
high nonlinearity \cite%
{ANW,JR,Battye97,Sutcliffe}.

A kink-like analytic solution was given by \cite{Sutcliffe}
\begin{equation}
F_{1}(x)=4\arctan [\exp (-x)],  \label{kink}
\end{equation}
with $E^{SK}=1.24035(6\pi ^{2}f_{\pi }/e)$, while an alternative
Skyrmion
profile, proposed based on the instanton method, takes the form\cite%
{Atiyah89}
\begin{equation}
F_{2}(x)=\pi \left[ 1-(1+\frac{\lambda }{x^{2}})^{-1/2}\right] ,
\label{F2}
\end{equation}
with corresponding energy $1.2432(6\pi ^{2}f_{\pi }/e)$ for the
numeric factor $\lambda =2.109$. The singularity at $r=0$ is gauge
dependent and can be gauged away without affecting the value for the
Skyrme field.

To find the more accurate analytic solution, we first improve the solution (%
\ref{kink}) into $4\arctan [\exp (-cx)]$ with $c$ a numeric factor
and then take $\lambda $ to be a $x$-dependent function: $\lambda
\rightarrow \lambda (x)$. Hence, we propose a Skyrmion profile
function in the hybrid form mixing (\ref{kink}) and (\ref{F2})
\begin{equation}
F_{w}(x)=4w\arctan [\exp (-cx)]+\pi (1-w)\left[ 1-(1+\frac{\lambda (x)}{x^{2}%
})^{-1/2}\right]   \label{Fw}
\end{equation}
with $w\in \lbrack 0,1]\,$\ being a positive weight factor. In
principle, one can find the governing equation for the unknown
$\lambda (x)$ by
substituting (\ref{Fw}) into (\ref{EOM1}) and obtain a series solution of $%
\lambda (x)$ by solving the governing equation. Here, however, we
choose the
Pad\'{e} approximation to parameterize $\lambda (x)$%
\begin{equation}
\lambda (x)=\lambda _{0}\frac{1+ax^{2}+\cdot \cdot \cdot
}{1+bx^{2}+\cdot \cdot \cdot } ,  \label{Pade}
\end{equation}
since it has as equal potential as series in approximating a
continuous
function. Note that we have already written $\lambda (x)$ as function of $%
x^{2}$ instead of $x$ since so is $F_{2}(x)$ in (\ref{F2}). The
simplest nontrivial case of the above Pad\'{e} approximation is the
[2/2] approximant
\begin{equation}
\lambda (x)=\lambda _{0}\frac{1+ax^{2}}{1+bx^{2}}.  \label{22}
\end{equation}

The minimization of the energy (\ref{Eden}) with the trial function (\ref{Fw}%
) with respect to the variational parameters ($a,b,c,w,\lambda
_{0}$) was carried out numerically for the [2/2] Pad\'{e}
approximant (\ref{22}) using the downhill simplex method (the
Neilder-Mead algorithm). The result for the
optimized parameters is given by%
\begin{equation}
\begin{array}{c}
a=0.330218,b=1.331975,c=2.094056, \\
w=0.286566,\lambda _{0}=7.323877.%
\end{array}
\label{para}
\end{equation}
with $E^{SK}=1.23152(6\pi ^{2}f_{\pi }/e)$. The solution (\ref{Fw}), with $%
\lambda (x)$ given by (\ref{Pade}) and the parameters (\ref{para}),
is referred as solution Hyb(2/2) for short in this paper and is
plotted in \textrm{Fig.1}, compared to the solutions (\ref{kink})
and (\ref{F2}), and the numerical solution (Num.) to the equation
(\ref{EOM1}). We also include the analytic solutions given by
\cite{Yamashita} and the solution in the form of the purely Pad\'{e}
approximant \cite{Ponciano} for comparison. A quite well agreement
of our solution with the numerical solution can be seen from this
plot. We note that the inequality $E^{SK}/(6\pi ^{2}f_{\pi }/e)\geq
|B|=1$ is fulfilled for all of cited results of the energy(see
\textrm{Table I}).

To check how well the asymptotic behavior of (\ref{Fw}) is we apply
the asymptotic expansion analysis on the profile $F(x)$. For small
$x\ll 1$ the solution $F(x)$ to the equation (\ref{EOM1}) is given
by
\begin{equation}
\begin{split}
F(x)=&\pi +Ax+Bx^{3}+Cx^{5}+\cdots\\
=&\pi -2.007528x+0.358987x^{3}-0.146499x^{5}+\cdots ,
\end{split}
\label{Fsmall}
\end{equation}%
(see also \cite{JMP}, where the variable $x$ used is twice of $x$ in
this paper) while the analytic solution (\ref{Fw}), when (\ref{22})
and (\ref{para}) is used, behaves like
\begin{equation}
F_{w}(x)=\pi -2.028368x+0.518855x^{3}-0.641539x^{5}+\cdots .
\label{Fwsmall}
\end{equation}%
One can see that (\ref{Fwsmall}) agrees well with (\ref{Fsmall}) up
to $x^{6} $. For large $x\rightarrow \infty $ the series solution
for $F(x)$ can be obtained by solving (\ref{EOM1}) with $x$ replaced
by $1/y$ and using the
series expansion for small $y$. After re-changing $y$ to $x$, one finds%
\begin{equation}
F(x)=\frac{2.1596}{x^{2}}\left\{ 1-\frac{0.222}{x^{4}}-\frac{116.0}{x^{6}}%
\right.
\left. +\frac{0.113}{x^{8}}+\frac{2.71}{x^{10}}+\cdots \right\} .%
\label{Flarge}
\end{equation}%
On the other hand, the solution Hyb(2/2), at large $x$, has the
asymptotic form
\begin{equation}
\begin{split}
F_{w}(x)=\frac{2.0348}{x^{2}}\left\{ 1+\frac{0.91576}{x^{2}}-\frac{5.8524}{%
x^{4}}+\frac{9.6819}{x^{6}}\right.  \\
\left. +\frac{3.1677}{x^{8}}-\frac{51.943}{x^{10}}%
+\cdots \right\} ,%
\end{split}
\label{Fla}
\end{equation}%
which agrees globally with (\ref{Flarge}) except for a small bit
differences. The detailed differences between (\ref{Flarge}) and
(\ref{Fla}) at large $x$ can be due to the fact that the
variationally-obtained solution (\ref{Fw}) approximates the Skyrmion
profile globally and may produce small
errors in local region, for instance, $F_{w}(50)=8.142\times 10^{-4}$ while $%
F(50)=8.638\times 10^{-4}$.

The disagreement can be improved by employing the Pad\'{e}
approximant of higher order than (\ref{22}), for example, the [4/4]
approximant
\begin{equation}
\lambda (x)=\lambda
_{0}\frac{1+ax^{2}+a_{4}x^{4}}{1+bx^{2}+b_{4}x^{4}}. \label{44}
\end{equation}
The minimization of (\ref{Eden}) using (\ref{44}), as done for the
[2/2] approximant, yields the numerically optimal parameters,
\begin{equation}
\begin{array}{c}
a=0.2598,b=0.5446,c=1.9932,a_{4}=0.0538, \\
b_{4}=0.1226,w=0.1839,\lambda _{0}=3.9439,%
\end{array}
\label{p2}
\end{equation}
The solution (\ref{Fw}) with $\lambda (x)$ specified by (\ref{44}) and (\ref%
{p2}) will be referred as the Hyb(4/4) in this paper and is also
plotted in \textrm{Fig.1}.\ The \textrm{Fig.2} shows the profiles of
$F(x)$ at large $x$ for Hyb(2/2) as well as Hyb(4/4), and the
numeric solution. The asymptotic expansion of the solution Hyb(4/4)
shows that for small $x\ll 1$ the profile becomes %
\[
F_{w}(x)=\pi -2.0243x+0.4654x^{3}-0.4270x^{5}+\cdots .\newline
\]
while for $x\rightarrow \infty $%
\begin{equation*}
\begin{split}
F_{w}(x)=\frac{2.220}{x^{2}}\left\{ 1-\frac{0.9150}{x^{2}}-\frac{9.5908}{%
x^{4}}-\frac{65.358}{x^{6}}\right.\\
\left. +\frac{264.78}{x^{8}}-\frac{821.88}{x^{10}}%
+\cdots \right\} ,
\end{split}
\end{equation*}
Here, a better value $F_{w}(50)=8.877\times 10^{-4}$ is obtained for
the latter asymptotic profile in contrast with the solution
Hyb(2/2). The computed Skyrmion energies (\ref{Eden}), measured in
the unit of $2f_{\pi }/e $, are listed in \textrm{Table I},
including the corresponding results obtained by the numeric solution
and obtained in the relevant references as indicated.

In solving (\ref{EOM1}) numerically, we employ the nonlinear shoot
algorithm for the boundary values at $x=0.001$ and $x=1000$ based on
the asymptotic formulas (\ref{Fsmall}) and (\ref{Flarge}) of the
chiral angle $F(x)$.

\section{The static properties of nucleons at low energy}

The static properties of nucleons can be extracted by
semi-classically quantizing the spinning modes of Skyrme Lagrangian
using the collective variables\cite{ANW}. Here, we will use the
solution Hyb(2/2) and Hyb(4/4) to compute the static properties of
nucleons and nucleon-isobar($\Delta $) within the framework of the
bosonic quantization of Skyrme model.

Following Adkin et al.\cite{ANW}, one can choose $SU(2)$-variable
$A(t)$ as the collective variables, and substitute
$U=A(t)U_{0}(x)A(t)^{\dagger }$
into (\ref{SK}). In the adiabatic limit, one has%
\begin{equation}
S^{SK}=S_{0}^{SK}+I_{0}\Lambda \int dtTr[\frac{\partial A}{\partial t}\frac{%
\partial A^{\dag }}{\partial t}],  \label{SS}
\end{equation}%
with $S_{0}^{SK}$ the action for the static hedgehog configuration, $%
I_{0}=\pi /(3e^{3}f_{\pi })$, and
\begin{equation}
\Lambda =8\int_{0}^{\infty }x^{2}dx\sin ^{2}F[1+F_{x}^{2}+\sin
^{2}F/x^{2}]. \label{lam}
\end{equation}%
which is independent of $f_{\pi }$ and $e$. The Hamiltonian associated to (%
\ref{SS}), when quantized via the quantization procedure in terms of
collective coordinates, yields an eigenvalue $\langle H\rangle
=M+J(J+1)/(2I_{0}\Lambda )$, with $M=E^{SK}$ being the soliton
energy of the Skyrmion. This yields the masses of the nucleon and
$\Delta $-isobar
\begin{equation}
M_{N}=M+\frac{3}{8I_{0}\Lambda },M_{\Delta
}=M+\frac{15}{8I_{0}\Lambda }. \label{ndm}
\end{equation}
By adjusting $f_{\pi }$ and $e$ to fit the nucleon and delta masses in (\ref%
{ndm}), one can fix the model parameters $f_{\pi }$ and $e$ using
the calculated $M$ and $\Lambda $ through (\ref{Eden}) and
(\ref{lam}).

The isoscalor root mean square(r.m.s) radius and isoscalor magnetic
r.m.s radius are given by
\begin{equation}
\begin{split}
ef_{\pi }\langle r^{2}\rangle _{I=0}^{1/2} =&\{-\frac{2}{\pi
}x^{2}\sin^{2}FF_{x}\}^{1/2}  \label{msr} \\
ef_{\pi }\langle r^{2}\rangle _{M,I=0}^{1/2} =&\left\{ \frac{%
\int_{0}^{\infty }x^{4}\sin ^{2}FF_{x}dx}{\int_{0}^{\infty
}x^{2}\sin ^{2}FF_{x}dx}\right\} ^{1/2}
\end{split}
\end{equation}
respectively. Combining with the masses of nucleon and the delta,
one can evaluate the magnetic moments for proton and neutron via the
following formula
\begin{equation}
\mu _{p,n}=\mu _{p,n}^{I=0}+\mu _{p,n}^{I=1}=\frac{\langle
r^{2}\rangle _{I=0}}{9}M_{N}(M_{\Delta }-M_{N})\pm
\frac{M_{N}}{2(M_{\Delta }-M_{N})}, \label{mu}
\end{equation}
where plus and minus correspond to proton and neutron, respectively.
The calculated results for these quantities using two solution
schemes (Hyb.(2/2) and Hyb.(4/4)) are shown explicitly in
\textrm{Table II}, compared to the experimental values as well as
that computed by the numeric solution for $F(x)$. The corresponding
results from other predictions are also shown in this table. Here in
\textrm{Table II}, we use the experimental values $M_{N}=938.9MeV$,
$M_{\Delta }=1232MeV$ for fixing $e$ and $f_{\pi }$ through
(\ref{ndm}), in contrast with the input $M_{N}=938MeV$, $M_{\Delta
}=1232MeV$ used by Ref.\cite{ANW} and Ref.\cite{Yamashita}.

To check the solution further, we also list, in the \textrm{Table
II}, the axial coupling constant and the $\pi NN$-coupling, which
are given by
\begin{equation}
g_{A}=-\frac{\pi }{3e^{2}}G , g_{\pi NN}=\frac{M_{N}}{f_{\pi }}g_{A}
\label{coup}
\end{equation}
respectively. Here, the numeric factor $G$ is
\begin{equation}
\begin{split}
G=4\int_{0}^{\infty }dxx^{2}\left[ F_{x}+\frac{\sin 2F}{x}+\frac{\sin 2F}{x}%
(F_{x})^{2}\right.  \\
\left. +\frac{2\sin ^{2}F}{x^{2}}F_{x}+\frac{\sin ^{2}F}{x^{3}}\sin
2F\right] .
\end{split}
\label{G}
\end{equation}

\section{Concluding remarks}

We show that the hybrid form of a kink-like solution and that given
by the instanton method are suited to approximate the exact solution
for the hedgehog Skyrmion, when combining with Pad\'{e}
approximation. The resulted analytic solution (\ref{Fw}) has two
remarkable features: (1) it is simple in the sense that it is
globally given in whole region; (2) it well approaches the
asymptotic behavior of the exact solution. We note that the
further generalization of (\ref{Fw}), made by approximating $c$ in (\ref{Fw}%
) via Pad\'{e} approximation, does not exhibit remarkable
improvement, particularly in the asymptotic behavior of the chiral
angle $F(x)$ at infinity. We expect that our solution can be useful
in the dynamics study of the Skyrmion evolution and interactions.

\section*{Acknowledgements}

D. J thanks C. Liu and ChuengRyong Ji for discussions. This work is
supported in part by the National Natural Science Foundation of
China (No.10547009) and (No.10965005), and the Knowledge and S\&T
Innovation Engineering Project of NWNU (No. NWNU-KJCXGC-03-41)

\begin{table*}
\begin{center}
\tabcolsep0.14in
\begin{tabular}{|c|c|c|c|c|c|c|c|c|}
\multicolumn{9}{c}{Table I}\\
\hline Work & \cite{ANW} & \cite{Ponciano} & \cite{Battye97} &
\cite{Yamashita} & \cite{JMP} & Hyb(2/2) & Hyb(4/4) & Num. \\
\hline$\frac{E^{SK}}{(2f_{\pi }/e)}$ & $36.5$ & $\allowbreak 36.47$ & $36.484$ & $%
\allowbreak 36.\allowbreak 474$ & $36.47$ & $36.4638$ & $36.4638$ &$36.4616$\\
\hline $\frac{M}{(6\pi ^{2}f_{\pi }/e)}$ & $1.233$ & $1.2317$ &
$1.2322$ & $1.23186$ & $1.2317$ & $1.23152$ & $1.23146$ & $1.23145$
\\ \hline
\end{tabular}
\end{center}
\end{table*}
\begin{table*}
\begin{center}
\tabcolsep0.2in
\begin{tabular}{|c|c|c|c|c|c|c|}
\multicolumn{7}{c}{Table II}\\
\hline Quantities & Ref.\cite{ANW} & Ref.\cite{Yamashita} &
Hyb.(2/2) & Hyb.(4/4) & Num. & $\text{Expt.}$ \\
\hline$
M/(2f_{\pi }/e)$ & $36.5$ & $36.5$ & $36.4638$ & $36.4638$ & $36.4616$ & $%
\mathbf{-}$ \\
\hline $2f_{\pi }(MeV)$ & $129$ & $130$ & $128.730$ & $129.453$ &
$129.260$ & $186$\\
\hline
$e$ & $5.45$ & $5.48$ & $5.4229$ & $5.4527$ & $5.4446$ & $-$ \\
\hline
$\Lambda $ & $50.9$ & $52.2$ & $50.1467$ & $51.2830$ & $50.9782$ & $-$ \\
\hline
$\langle r^{2}\rangle _{I=0}^{1/2}(fm)$ & $0.59$ & $0.586$ & $0.5985$ & $%
0.5920$ & $0.5938$ & $0.72$ \\
\hline$\langle r^{2}\rangle _{M,I=0}^{1/2}(fm)$ & $0.92$ & $0.920$ & $0.9258$ & $%
0.9208$ & $0.9222$ & $0.81$ \\ \hline $\mu _{p}$ & $1.87$ & $-$ &
$1.8825$ & $1.8764$ & $1.8781$ & $2.79$ \\ \hline
$\mu _{n}$ & $-1.31$ & $-$ & $-1.3209$ & $-1.3269$ & $-1.3253$ & $-1.91$ \\
\hline
$|\mu _{p}/\mu _{n}|$ & $1.43$ & $-$ & $1.4252$ & $1.4141$ & $1.4171$ & $%
1.46 $ \\ \hline $g_{A}$ & $0.61$ & $-$ & $0.6332$ & $0.5992$ &
$0.6081$ & $1.23$ \\ \hline
$g_{\pi NN}$ & $8.9$ & $-$ & $9.2364$ & $8.6921$ & $8.8343$ & $13.5$ \\
\hline
\end{tabular}
\end{center}
\end{table*}
\begin{figure}
\begin{center}
\rotatebox{0}{\resizebox *{8.5cm}{7.7cm} {\includegraphics
{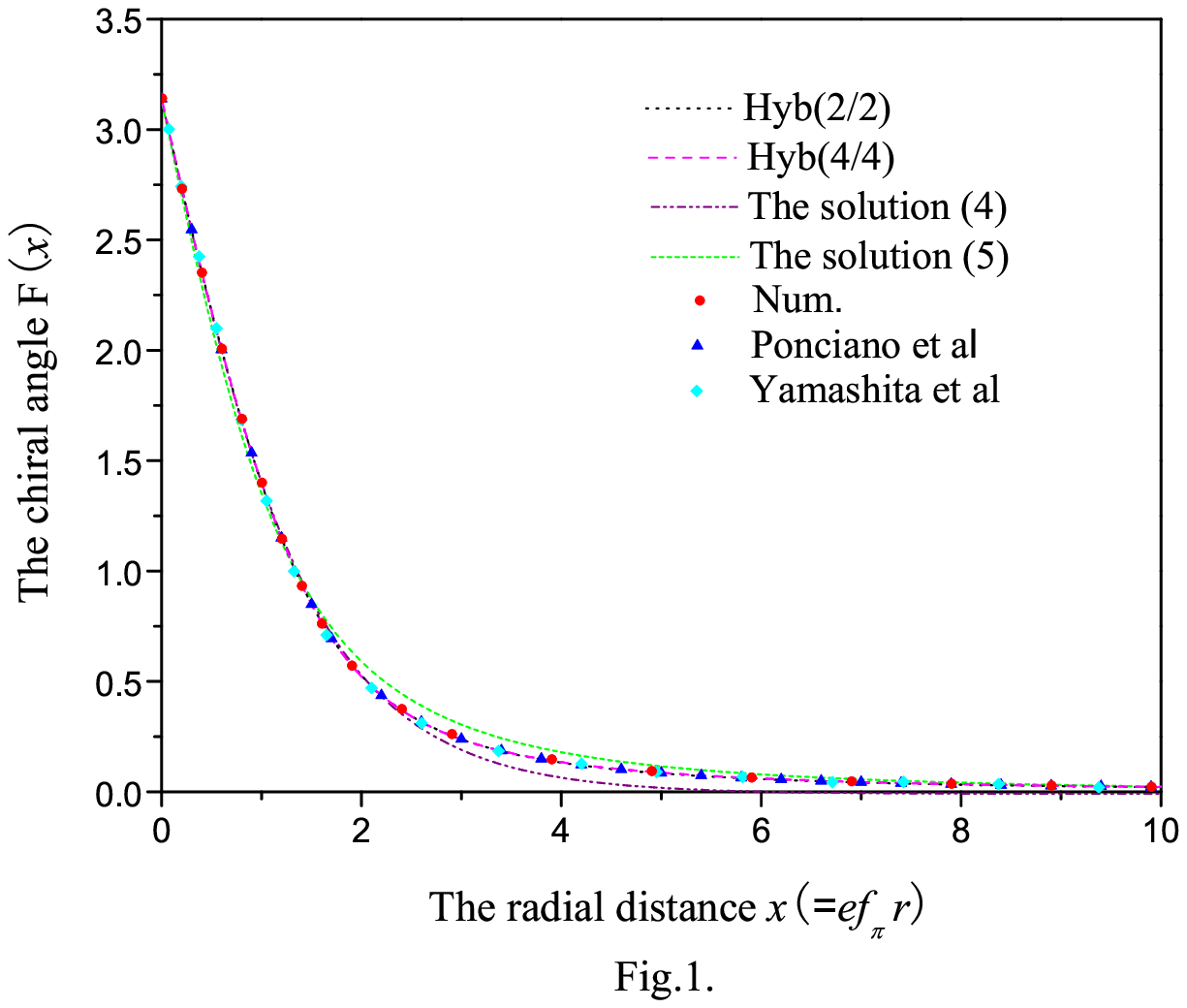}}}
\end{center}
\end{figure}
\begin{figure}
\begin{center}
\rotatebox{0}{\resizebox *{8.5cm}{7.5cm} {\includegraphics
{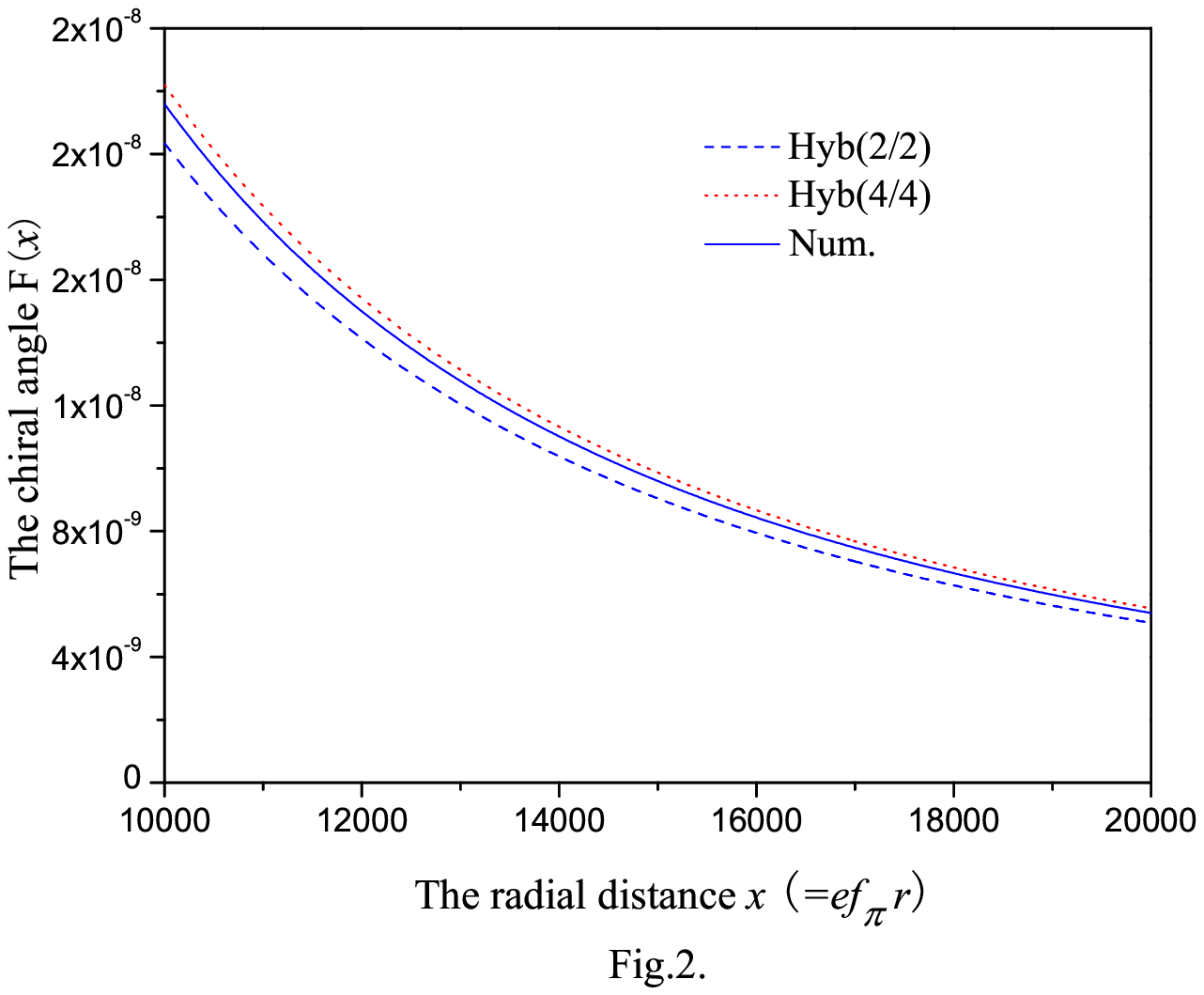}}}
\end{center}
\end{figure}


\section*{References}

\begin{thebibliography}{99}
\bibitem{Skyrme} T.H.R. Skyrme, Nucl. Phys. 31(1961) 556.
\bibitem{Witten} E. Witten, Nucl. Phys. B 223(1983) 422.
\bibitem{ANW} G.S. Adkins, C.R. Nappi and E. Witten, Nucl. Phys. B 228(1983) 552.
\bibitem{JR} A. Jackson, A.D. Jackson et al., Nucl. Phys. A 432(1985) 567.
\bibitem{Sutcliffe} P.M. Sutcliffe, Phys. Lett. B 292(1992) 104.
\bibitem{Battye97} R. Battye and P.M. Sutcliffe, Phys. Rev. Lett.79(1997) 363.
\bibitem{AtiyahManton} M.F. Atiyah and N.S. Manton, Commun. Math. Phys.152(1993) 391.
\bibitem{Atiyah89} M.F. Atiyah and N.S. Manton, Phys. Lett. B 222(1989) 438.
\bibitem{Battye} R. Battye and P.M. Sutcliffe, Phys. Rev. C 73(2006) 055205.
\bibitem{Zahed} I. Zahed\ and G. Brown, Phys. Rept. 142(1986) 1.
\bibitem{Topsoliton} N.S. Manton and P.M. Sutcliffe, Topological Solitons,Cambridge Univ. Press, Cambridge, 2004.
\bibitem{JMP} J. Ananias et al., J. Math. Phys. 32, 7(1991) 1949.
\bibitem{Ponciano} J.A. Ponciano et al., Phys. Rev. C 64(2001) 045205.
\bibitem{Yamashita} J. Yamashita and M. Hirayama, Phys. Lett. B 642(2006) 160.
\bibitem{Faddeev} L.D. Faddeev, Lett. Math. Phys. 1(1976) 289.
\end{thebibliography}
\end{document}